\documentstyle[aps,prl,epsfig,twocolumn,graphics]{revtex}

\def\be{\begin{equation}}
\def\ee{\end{equation}}
\def\bea{\begin{eqnarray}}
\def\eea{\end{eqnarray}}

\begin{document}
\draft

\title{Fast quantum gates for neutral atoms}

\author{D.~Jaksch, J.I.~Cirac, and P.~Zoller}

\address{Institut f\"ur Theoretische Physik, Universit\"at Innsbruck,
Technikerstrasse 25, A--6020 Innsbruck, Austria.}

\author{S.L.~Rolston}

\address{National Institute of Standards and Technology, Gaithersburg, Maryland 20899.}

\author{R.~C\^{o}t\'{e}$^{1}$ and M.D.~Lukin$^{2}$}
\address{$^{1}$Physics Department, University of Connecticut, 2152 Hillside Rd., Storrs, Connecticut 06269-3046. \\
$^{2}$ ITAMP, Harvard-Smithsonian Center for Astrophysics, Cambridge, 
MA 02138.}

\date{\today}

\maketitle

\begin{abstract}
We propose several schemes for implementing a fast two-qubit quantum gate for neutral atoms with the gate operation time much faster than the time scales associated with the external motion of the atoms in the trapping potential. In our example, the large interaction energy required to perform fast gate operations is provided by the dipole-dipole interaction of atoms excited to low-lying Rydberg states in constant electric fields. A detailed analysis of imperfections of the gate operation is given.
\end{abstract}


\narrowtext

In recent years, numerous proposals to build quantum information processors have been made \cite{overview}. Due to their exceptional ability of quantum control and long coherence times, quantum optical systems such as trapped ions \cite{ions} and atoms \cite{trapping}, and cavity QED \cite{qed}, have taken a leading role in implementing quantum logic in the laboratory. {\em Quantum computing with neutral atoms} \cite{cc} seems particularly attractive in view of very long coherence times of the internal atomic states and well-developed techniques for cooling and trapping atoms in optical lattices, far off-resonance light traps and magnetic microtraps \cite{trapping}. Preparation and rotations of single qubits associated with long-lived internal states can be performed by addressing individual atoms with laser pulses. A central issue is to design {\em fast} two-qubit gates. 

First of all, it is difficult to identify a {\em strong} and controllable two-body interaction for neutral atoms, which is required to design a gate. Furthermore, the strength of two-body interactions does not necessarily translate into a useful fast quantum gate: large interactions are usually associated with strong mechanical forces on the trapped atoms. Thus, internal states of the trapped atoms (the qubits) may become entangled with the motional degrees of freedom during the gate, resulting effectively in an additional source of decoherence. This leads to the typical requirement that the process is  {\em adiabatic} on the time scale of the oscillation period of the trapped atoms in order to avoid entanglement with motional states. As a result, extremely tight traps and low temperatures are required. 

In the present Letter we propose a fast phase gate for neutral trapped atoms, corresponding to a truth table $|\epsilon_1\rangle \otimes |\epsilon_2\rangle \rightarrow e^{i \epsilon_1 \epsilon_2 \varphi} |\epsilon_1\rangle \otimes |\epsilon_2\rangle $ for the logical states $|\epsilon_i\rangle$ with $\epsilon_i=0,1$, which (i) exploits the very large interactions of permanent dipole moments of laser excited Rydberg states in a constant electric field to entangle atoms, while (ii) allowing gate operation times set by the time scale of the laser excitation or the two particle interaction energy, which can be significantly shorter than the trap period. Among the attractive features of the gate are the insensitivity to the temperature of the atoms and to the variations in atom-atom separation.

Rydberg states \cite{Gallagher} of a hydrogen atom within a given manifold of a fixed principal quantum number $n$ are degenerate. This degeneracy is removed by applying a constant electric field $\cal E$ along the $z$-axis (linear Stark effect). For electric fields below the Ingris-Teller limit the mixing of adjacent $n$-manifolds can be neglected, and the energy levels are split according to $\Delta E_{nqm}= 3 n q e a_0 {\cal E} /2$ with parabolic and magnetic quantum numbers $q=n-1-|m|,n-3-|m|,\ldots,-(n-1-|m|)$ and $m$, respectively, $e$ the electron charge, and $a_0$ the Bohr radius. These Stark states have permanent dipole moments ${\bf \mu} \equiv\mu_z {\bf e}_z = 3 n q e a_0 {\bf e}_z/2$. In alkali atoms the $s$ and $p$-states are shifted relative to the higher angular momentum states due to their quantum defects, and the Stark maps of the $m=0$ and $m=1$ manifolds are correspondingly modified \cite{Gallagher}.

\begin{figure}[tbp]
\includegraphics[width=8cm]{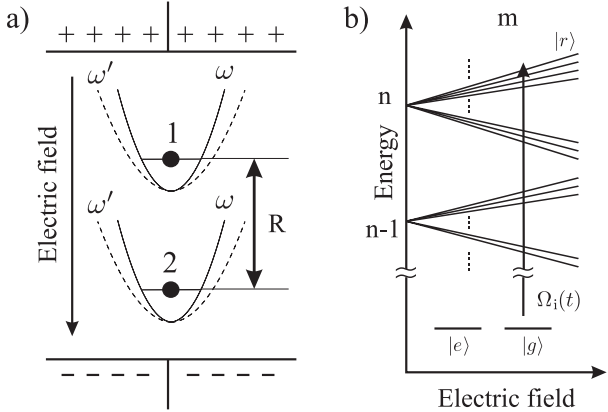}
\caption{
(a) Setup: A constant electric field along the $z$ direction is applied to alkali atoms trapped in micro-traps. (b) Level scheme: Two ground states $|g\rangle$ and $|e\rangle$ (qubits), and laser excitation to the Rydberg state  $|g\rangle\rightarrow|r\rangle$.
} 
\label{levelscheme}
\end{figure}

Consider two atoms $1$ and $2$ at fixed positions (see Fig.~\ref{levelscheme}a), and initially prepared in Stark eigenstates, with a dipole moment along $z$ and a given $m$, as selected by the polarization of the laser exciting the Rydberg states from the ground state. They interact and evolve according to the dipole-dipole potential
\be
V_{\rm dip}({\bf r})= \frac{1}{4 \pi \epsilon_0} \left[
\frac{{\bf \mu}_1 \cdot {\bf \mu}_2}{|{\bf r}|^3}-
3 \frac{({\bf \mu}_1 \cdot {\bf r})({\bf \mu}_2 \cdot {\bf r})}{|{\bf r}|^5} \right],
\ee
with ${\bf r}$ the distance between the atoms. We are interested in the limit where the electric field is sufficiently large so that the energy splitting between two adjacent Stark states is much larger than the dipole-dipole interaction. For two atoms in the given initial Stark eigenstate, the diagonal terms of $V_{\rm dip}$ provide an energy shift whereas the non-diagonal terms couple $(m,m)\rightarrow(m\pm 1,m\mp 1)$ adjacent $m$ manifolds with each other. We will assume that these transitions are suppressed by an appropriate choice of the initial Stark eigenstate \cite{footnote}. As an illustration we choose the hydrogen state $|r\rangle=|n,q=n-1,m=0\rangle$ and find for a fixed distance ${\bf r}=R {\bf e}_z$ of the two atoms that $u(R)=\langle r| \otimes \langle r | V_{\rm dip}(R {\bf e}_z) | r \rangle \otimes |r \rangle$ is equal to $u(R)=-9 [n(n-1)]^2 (a_0/R)^3(e^2/8 \pi \epsilon_0 a_0)\propto n^4$. In alkali atoms we have to replace $n$ by the effective quantum number $\nu$ \cite{Gallagher}. We will use this large energy shift to entangle atoms.

The configuration we have  in mind is as follows (see Fig.~\ref{levelscheme}). We consider two atoms, which for the moment are assumed to be at fixed positions ${\bf x}_{j}$  with $j=1,2$ labeling the atoms, at a distance $R=|{\bf x}_1-{\bf x}_2|$. We store qubits in two internal atomic ground states (e.g.~hyperfine levels) denoted by $|g\rangle_j  \equiv |0\rangle_j$ and $|e\rangle_j  \equiv |1\rangle_j$. The ground states $|g\rangle_j$ are coupled by a laser to a given Stark eigenstate $|r\rangle_j$. The internal dynamics is described by a model Hamiltonian
\bea
&&H^i(t,{\bf x}_1,{\bf x}_2)= u |r\rangle_1 \langle r|\otimes |r\rangle_2 \langle r| \\
&&\quad + \sum_{j=1,2} \left[(\delta_j(t)- i \gamma) |r\rangle_j \langle r|
- \frac{\Omega_j(t,{\bf x}_j)}{2} \left(|g\rangle_j \langle r|+\rm{h.c.} \right)\right], \nonumber 
\eea
with $\Omega_j(t,{\bf x}_j)$  Rabi frequencies, and $\delta_j(t)$ detunings of the exciting lasers. $\gamma$ accounts for loss from the excited states $|r\rangle_j$.

Including the atomic motion, the complete Hamiltonian has the structure
\begin{mathletters}
\bea \label{Ha}
H(t,{\bf \hat x}_1, {\bf \hat x}_2) &=& H^T({\bf \hat x}_1, {\bf \hat x}_2) + H^i(t,{\bf \hat x}_1, {\bf \hat x}_2) \\ \label{Hb}
&\equiv& H^e(t,{\bf \hat x}_1, {\bf \hat x}_2) + H^i(t,{\bf x}_1,{\bf x}_2),
\eea
\end{mathletters}
where $H^T$ describes the motion and trapping of the atoms, and ${\bf \hat x}_j$ are the atomic position operators, and define ${\bf \hat r}={\bf \hat x_1}-{\bf \hat x_2}$. Our goal is to design a phase gate for the internal states with a gate operation time $\Delta t$ with the internal Hamiltonian $H^i(t,{\bf x}_1,{\bf x}_2)$ in Eq.~(\ref{Hb}), where (the c--numbers) ${\bf x}_j$ now denote the centers of the initial atomic wave functions as determined by the trap, while avoiding motional effects arising from $H^e(t,{\bf \hat x}_1, {\bf \hat x}_2)$. This requires that the gate operation time $\Delta t$ is short compared to the typical time of evolution of the external degrees of freedom, $H^e \Delta t \ll 1$. Under this condition, the initial density operator of the two atoms evolves as $\rho_e \otimes \rho_i \rightarrow \rho_e \otimes \rho_i'$ during the gate operation. Thus the motion described by $\rho_e$ does not become entangled with the internal degrees of freedom given by $\rho_i$. Typically, the Hamiltonian $H^T$ will be of the form
\bea \label{HT}
H^T&=&\sum_{j=1,2} \left[ \left( \frac{{\bf \hat p}_j^2}{2m} + V^T_j({\bf \hat x}_j) \right) \left(|g\rangle_j \langle g| +|e\rangle_j \langle e| \right) + \right. \nonumber \\ 
&&\qquad \left. \left( \frac{{\bf \hat p}_j^2}{2m} + V^r_j({\bf \hat x}_j) \right) |r\rangle_j \langle r| \right],
\eea
which is the sum of the kinetic energies of the atoms and the trapping potentials for the various internal states. 
In our estimates for the effects of motion we will assume that the potentials are harmonic with a frequency $\omega$ for the ground states, and $\omega'$ for the excited state.

Physically, for the splitting of the Hamiltonian according to Eq.~(\ref{Hb}) to be meaningful we require the initial width of the atomic wave function $a$, as determined by the trap, to be much smaller than the mean separation between the atoms $R$. We expand the dipole-dipole interaction around $R$, $V_{\rm dip}(\hat {\bf r})=u(R)-F ({\bf \hat r}_z-R)+\ldots$, with $F=3 u(R)/R$. Here the first term gives the energy shift if both atoms are excited to state $|r\rangle$, while the second term contributes to $H^e$ and describes the mechanical force on the atoms due to $V_{\rm dip}$. Other contributions to $H^e$ arise from the photon kick in the absorption $|g\rangle\rightarrow|r\rangle$, but these terms can be suppressed in a Doppler-free two photon absorption, for example.
We obtain $ H^e({\bf \hat x}_1, {\bf \hat x}_2)=H^T-F({\bf \hat r}_z-R) |r\rangle_1 \langle r|\otimes |r\rangle_2 \langle r|. $

\begin{figure}[tbp]
\includegraphics[width=8cm]{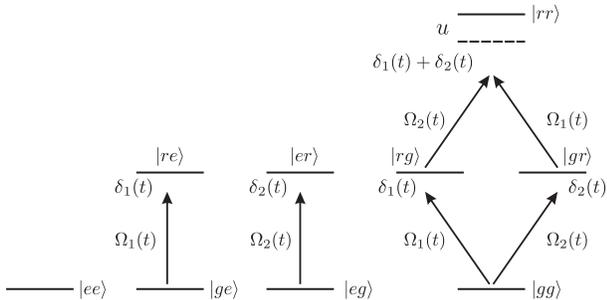}
\caption{Schematics of the ideal scheme. The internal state $|g\rangle_j$ is coupled to the excited state $|r\rangle_j$ by the Rabi frequency $\Omega_j(t)$ with the detuning $\delta_j(t)$. The state $|e \rangle_j$ decouples from the evolution of the rest of the system.}
\label{fastgate}
\end{figure}

We will now study several models for phase gates according to dynamics induced by $H^i$. A schematic overview of the internal evolution of the two Rydberg atoms is given in Fig.~\ref{fastgate} (with shorthand notation $|g\rangle_1 \otimes |e\rangle_2 \equiv |ge\rangle$ etc.).

{\em Model A:} We assume $\Omega_j \gg u$, and in this scheme individual addressing of the atoms is not necessary, i.e., $\Omega_1=\Omega_2=\Omega$. We set $\delta_1=\delta_2=0$. We perform the gate with three steps: (i) apply a $\pi$-pulse to the two atoms, (ii) wait for a time $\Delta t=\varphi/u$, and (iii) apply again a $\pi$-pulse to the two atoms. Since the Rabi frequency $\Omega$ is much larger than the interaction energy the first pulse will transfer all the occupation from the states $|g\rangle_j$ to the states $|r \rangle_j$ and the second laser pulse will bring the population back to the ground states $|g\rangle_j$. Between the two pulses the state $|rr\rangle$ will pick up the extra phase $\varphi=u \Delta t$. Thus, this scheme realizes a phase gate operating on the time scale $\Delta t \propto 1/u$. We note that the accumulated phase depends on the precise value of $u$, i.e.\ is sensitive to the atomic distance. The probability of loss due to $\gamma$ is approximately given by $p_l=2 \varphi \gamma /u$. Furthermore, during the gate operation (i.e. when the the state $|rr\rangle$ is occupied) there are large mechanical effects due to the force $F$. This motivates the following model.

{\em Model B:} We assume $u \gg \Omega_j$. Let us for the moment assume that the two atoms can be addressed individually,\cite{addressing}, i.e.~$\Omega_1(t) \neq \Omega_2(t)$. We set $\delta_j=0$  and perform the gate operation in three steps: (i) We apply a $\pi$-pulse to the first atom, (ii) a $2 \pi$-pulse (in terms of the unperturbed states, i.e.~it has twice the pulse area of pulse applied in (i)) to the second atom, and, finally, (iii) a $\pi$-pulse to the first atom. As can be seen from Fig.~\ref{fastgate}, the state $|ee\rangle$ is not affected by the laser pulses. If the system is initially in one of the states $|ge\rangle$ or $|eg\rangle$ the pulse sequence (i)-(iii) will cause a sign change in the wave function. If the system is initially in the state $|gg\rangle$ the first pulse will bring the system to the state $i |rg\rangle$, the second pulse will be {\em detuned} from the state $|rr \rangle$ by the interaction strength $u$, and thus accumulate a {\em small} phase $\tilde \varphi \approx \pi \Omega_2 / 2 u \ll \pi$. The third pulse returns the system to the state $ e^{i (\pi- \tilde \varphi)} |gg\rangle$, which realizes a phase gate with $\varphi=\pi-\tilde \varphi \approx \pi$ (up to trivial single qubit phases). The time needed to perform the gate operation is of the order $\Delta t \approx 2\pi/ \Omega_1 + 2\pi/\Omega_2$. Loss from the excited states $|r\rangle_j$ is small provided $\gamma \Delta t \ll 1$, i.e.~$\Omega_j\gg \gamma$. If we choose $u=1.8 \rm{GHz}$, $\Omega_j=100 \rm{MHz}$ and $\gamma=100 \rm{kHz}$ \cite{params}
%corresponding to a typical decay rate for low lying Rydberg states, 
we find a probability of loss from the excited states of $p_l=3.4\%$.  

\begin{figure}[tbp]
\includegraphics[width=8cm]{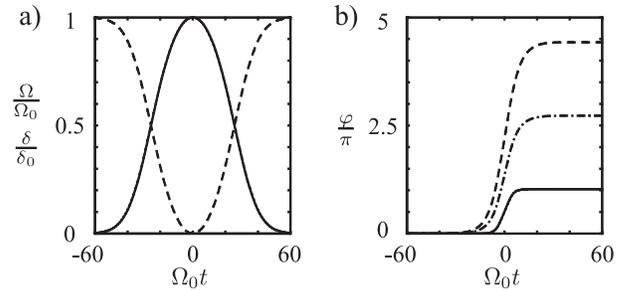}
\caption{a) The Rabi frequency $\Omega(t)$ normalized to $\Omega_0=100\rm{MHz}$ (solid curve), $\delta(t)$ normalized to $\delta_0=1.7\rm{GHz}$ (dashed curve). We chose $\gamma=100\rm{kHz}$, and $u=1.8\rm{GHz}$. b) Accumulated phase of the state $|gg\rangle$ (dashed curve) and $|ge\rangle$ (dash-dotted curve). The resulting accumulated $\varphi(t)$ is given by the solid curve and the probability of loss from the excited state is found to be $p_l=1.7\%$.}
\label{phase}
\end{figure}

An {\em adiabatic} version of this gate has the advantage that individual addressing of the two atoms is not required, $\Omega_{1,2}(t)\equiv\Omega(t)$ and $\delta_{1,2}(t)\equiv\delta(t)$. In this scheme we assume the time variation of the laser pulses to be slow on the time scale given by $\Omega$ and $\delta$ (but still larger than the trap oscillation frequency), so that the system adiabatically follows the dressed states of the Hamiltonian $H^i$. After adiabatically eliminating the state $|rr\rangle$, we find the energy of the dressed level adiabatically connected to the initial state $|gg\rangle$ to be given by
$\epsilon_{gg}(t)={\rm sgn}(\tilde \delta)
\left(\left|\tilde \delta \right| - (\tilde 
\delta^2+2 \Omega^2)^{1/2} \right)/2
$  with $\tilde \delta = \delta -\Omega^2/(4 \delta + 2 u)$ the detuning including a Stark shift.
For the dressed levels connected to  $|eg\rangle$ and $|ge\rangle$ we have $\epsilon_{eg}(t)={\rm sgn}(\delta) \left(\left|\delta \right| - (\delta^2+ \Omega^2)^{1/2} \right)/2$. The entanglement phase follows as $\varphi(t)=\int_0^t dt' \left(\epsilon_{gg}(t') - 2 \epsilon_{eg}(t')\right)$.
For a specific choice of pulse duration and shape $\Omega(t)$ and $\delta(t)$ we achieve $\varphi\equiv\varphi(\Delta t)=\pi$ (see Fig.~\ref{phase}a). In Fig.~\ref{phase}b the phases accumulated in the dressed states of $|gg\rangle$ and  $|eg\rangle (|ge\rangle)$, and the resulting entanglement phase $\varphi$ are shown. To satisfy the adiabatic condition, the gate operation time $\Delta t$ is approximately one order of magnitude longer than in the gate discussed above. 

A remarkable feature of model B is that, in the ideal limit, the doubly excited state $|rr\rangle$ is never populated. Hence, the mechanical effects due to atom-atom interaction are greatly suppressed. Furthermore, this version of the gate is only weakly sensitive to the exact distance between the atoms, since the distance-dependent part of the entanglement phase $\tilde \varphi \ll \pi$ \cite{pillet}. These features allow one to design robust quantum gates with atoms in lattices that are {\it not} filled regularly.

We now turn to a discussion of decoherence mechanisms, which include spontaneous emission, transitions induced by black body radiation, ionization of the Rydberg states due to the trapping or exciting laser fields, and motional excitation of the trapped atoms. While dipole-dipole interaction increases with $\nu^4$, the  spontaneous emission and ionization of the Rydberg states by optical laser fields decreases proportional to $\nu^{-3}$. For $\nu < 20$  the black body radiation is negligible in comparison with spontaneous emission, and similar conclusions hold for typical ionization rates from the Rydberg states for the numbers quoted in the context of Fig.~\ref{phase}.

We now calculate the motional effects described by $H^e$ on the fidelity of the gate. The dipole-dipole force, given by $F$, causes a momentum kick to both atoms when they populate $|rr\rangle$. We assume the atoms to be initially in the ground state of the trapping potential. For the adiabatic gate (Model B), we estimate the probability $p_k$ of exciting a trap state without changing the internal state of the atoms in time dependent perturbation theory. We find that the perturbative transition probability is bounded by
$p_k <   \left(3 \eta \Omega_0^2 \Delta t/8 u \right)^2/2$ with $\eta=a/R \ll 1$. For the parameters given in Fig.~\ref{phase}, and $\eta=1/30$ we find numerically $p_k \approx 2.5 \times 10^{-7}$ while the analytic approximation with a gate operation time $\Delta t=100 /\Omega_0$ yields $p_k < 2.4 \times 10^{-3}$. The probabilities of exciting motional quanta, and at the same time changing the internal state of the atoms, require the perturbation to induce internal transitions with an energy difference of $\Omega$ or $u$ in addition to causing a motional excitation. Since $u \gg \Omega \gg \omega$ whenever there is population in the state $|rr\rangle$, these probabilities are much smaller than $p_k$, and are thus negligible. For finite temperature of the trapped atoms, we have to incoherently sum the probabilities of exciting an atom due to the kick for the different trap states. We find $p_k^{\bar n} =(2 \bar n +1) p_k$ where $\bar n$ is the mean excitation number in the trap as determined by the finite temperature $T$.

The optical potential seen by the atom in the Rydberg state $|r\rangle$ (see Eq.~(\ref{HT})) can be different from the trapping potential in the ground states $|g\rangle$, $|e\rangle$. This difference causes the motional wave function to change its shape while the atom is in $|r\rangle$, and thus excites the atomic motion. We treat the deviation of the potential in the excited state from the ground state trapping potential as a perturbation and estimate the probability $p_t$ of exciting an atom from the vibrational ground state during the gate operation. We find that the perturbation theory expression is bounded by 
$p_t< |\omega^2 - \omega'^2| \Delta t^2 / 128$.
For the parameters given in Fig.~\ref{phase}, and trap frequencies of $\omega=1{\rm MHz}$ and $\omega'=500{\rm kHz}$ we find numerically $p_t \approx 10^{-5}$. The analytical approximation (with  $\Delta t=100 /\Omega_0$) gives $p_t < 3.9 \times 10^{-3}$. At finite temperature $T$ we find by incoherent summation $p_t^{\bar n}=(2 \bar{n^2} + 2\bar n +1) p_t$. 

Alternatively, the optical trapping potential can be turned off for the short time of the gate operation. The shape of the wave function of the atoms evolves then independently of their internal states $|e\rangle$, $|g\rangle$ or $|r\rangle$. Therefore, no entanglement between external and internal degrees of freedom is created during the gate operation. However, the releasing and retrapping will cause heating of the atoms. We estimate this effect at finite temperature $T$ and find an increase in the mean excitation number of
$
\bar n \rightarrow \bar n + \Delta \bar n = \bar n + \left(\omega \Delta t\right)^2 (2 \bar n + 1) /4
$
during a gate operation.

The influence of imperfections on the other schemes discussed in this Letter can be estimated in the same way as for the adiabatic gate. However, in model A the momentum kick will be stronger than for the other schemes because all of the population is transferred to $|rr\rangle$ for a time $\varphi/u$. The force experienced by the atoms during this time is approximately given by $F$, yielding a total momentum transfer of $3 \varphi /R$ to the atoms, which for a separation $R \sim \lambda$ is of the order of several recoils $\hbar k$ with $k=2\pi / \lambda$. Therefore, only if the atoms are confined deep in the Lamb-Dicke regime $2 \pi a / \lambda \ll 1$ with $\lambda$ the optical wave length, will this momentum transfer not cause significant excitation.

DJ and PZ thank NIST Laser Cooling and Trapping Group for hospitality during their stay in Oct / Nov 1999, where this work was started. PZ thanks the quantum optics group at the University of Hannover for hospitality, and the Alexander von Humboldt foundation for support. RC and MDL acknowledge discussions with P.L. Gould and E.E. Eyler. Research at the University of Innsbruck is supported by the Austrian Science Foundation, and by the TMR network ERB-FMRX-CT96-0087. MDL is supported by NSF through ITAMP, and RC through NSF Grant No. PHY-9970757, and SLR through the ONR.


\begin{references}
\bibitem{params}
In Na these parameters can be achieved for $\nu=18$, $R=300{\rm nm}$ with an applied electric field of ${\cal E}=100{\rm V/cm}$, and a laser intensity of approximately $I=40{\rm kW/cm^2}$.
\end{references}
\end{document}